# Generalized Modified Blake-Zisserman Robust Spline Adaptive Filter for Generalized Gaussian Noise

**Haiquan Zhao, Senior Member, IEEE**

School of Electrical Engineering, Southwest Jiaotong University

**Bei Xu**

School of Electrical Engineering, Southwest Jiaotong University

*Abstract*—The spline adaptive filtering (SAF) algorithm-based information-theoretic learning has exhibited strong convergence performance in nonlinear system identification (NSI), establishing SAF as a promising framework for adaptive filtering. However, existing SAF-based methods suffer from performance degradation under generalized Gaussian noise (GGN) environment and exhibit significant steady-state misalignment under impulse noise. Moreover, prior research on SAF algorithms has not effectively addressed the adverse effects caused by outliers. To overcome these challenges, the generalized modified Blake-Zisserman robust spline adaptive filtering (SAF-GMBZ) algorithm is proposed. Compared to conventional SAF algorithms, SAF-GMBZ exhibits superior learning performance in GGN. Furthermore, the mean convergence ranges of the step-sizes and the steady-state mean-square error (MSE) are calculated by introducing the commonly utilized assumptions. To arrive at good convergence accuracy and noise cancellation capability in active noise control (ANC) application, the filter-c GMBZ (FcGMBZ) algorithm is further developed based on SAF-GMBZ. Simulation results confirm the accuracy of the theoretical steady-state MSE, and the superiority of the SAF-GMBZ algorithm under GGN environment in NSI, along with the effectiveness of the FcGMBZ algorithm in ANC application under impulsive noise environment.

This work was partially supported by National Natural Science Foundation of China (grant: 62171388, 61871461, 61571374).

Haiquan Zhao and Bei Xu are with the Key Laboratory of Magnetic Suspension Technology and Maglev Vehicle, Ministry of Education, Southwest Jiaotong University, Chengdu 610032, China, and also with the School of Electrical Engineering, Southwest Jiaotong University, Chengdu 610032, China (e-mail: hqzhao_swjtu@126.com; beixu_swjtu@126.com). (*Corresponding author: Haiquan Zhao*.)

## I. INTRODUCTION

With the advent of signal processing and the advancements in control systems, adaptive filtering techniques are widely regarded as a crucial research topic. These techniques have been intensively studied and widely used in different domains, including nonlinear system identification (NSI), acoustic echo cancellation (AEC), and active noise control (ANC) [1]. Among linear adaptive filtering algorithms (AFAs), several classic methods are widely recognized. The least-mean-square (LMS) algorithm is notable for its computational efficiency [2], the normalized LMS (NLMS) algorithm for its superior tracking performance [2], and the affine projection (AP) algorithm for its fast convergence speed [3]. However, despite their widespread use, these linear adaptive filtering algorithms face challenges such as high computational complexity and slow convergence in NSI. Therefore, nonlinear spline adaptive filters (SAF) [4] have emerged as an effective solution, offering significant advantages in adapting to signal variations due to their increased flexibility and robustness [5], [6], [7], [8]. The SAF algorithm significantly improves the nonlinear signal processing ability by employing the spline interpolation function as the nonlinear component of the filter. This approach provides high efficiency, reduced computational complexity, and improved adaptability.

Depending on the structural topology, SAF architectures are categorized into the Wiener [9] model, the Hammerstein [10] model, and multiple variants derived from the topology of these two models [11]. Over time, SAF has evolved through integration with conventional adaptive algorithms, improving performance. For instance, the SAF-LMS [4] algorithm was developed, with theoretical analysis accurately predicting its convergence behavior [12]. To improve the stability issue arising from eigenvalue spread in the input signal's autocorrelation matrix, a novel SAF algorithm leveraging the NLMS approach (SAF-NLMS) [13] was developed. Furthermore, Liu *et al.* incorporated a sign function into the NLMS algorithm (SNLMS) to propose the SAF-SNLMS [14], [15] algorithm, which mitigates the impact of impulse noise by minimizing the absolute value of the *a posteriori* error. Despite these advancements, existing SAF algorithms remain susceptible to non-Gaussian noise, limiting their effectiveness in adaptive filtering. Consequently, more advanced nonlinear SAF algorithms have been explored [16], [17], [18], [19].

Recently, information-theoretic learning-based AFAs have significantly improved performance in non-Gaussian noise environment. Among these, the maximum correntropy criterion (MCC) [20] has gained widespread adoption due to its effectiveness in NSI and ANC. Extending this approach, Peng *et al.* developed the SAF-MCC [21], [22] algorithm, which leverages the MCC criterion to suppress outliers and achieve superior performance in heavy-tailed non-Gaussian environments. This work has inspired further innovations aimed at enhancing SAF architectures. Notable advancements include the IIR-SAF-NLMM algorithm [23], which integrates normalized least M-estimation (NLMM) into an infinite impulse response (IIR)-based SAF framework for improved robustness, and the SAF-GMVC algorithm [24], which applies the generalized Versoria criterion (GMVC) to enhance performance in non-Gaussian environments. Particularly, the extension of the MCC criterion to the generalized MCC (GMCC) [25] function has led to the development of the SAF-GMCC algorithm. This extension improves robustness across various noise distributions, demonstrating superior performance compared to SAF-MCC in ANC application. However, the SAF-GMCC algorithm remains vulnerable to outliers in generalized Gaussian noise (GGN) environments and exhibits significant steady-state misalignment under impulse noise, necessitating further improvements.

In adaptive filtering, outliers refer to anomalous data points that substantially deviate from the expected distribution due to impulse noise, measurement errors, or sudden interferences. In generalized Gaussian environments, heavy-tailed noise, such as α-stable or Laplace noise, often produces extreme outliers, which can adversely affect convergence and introduce estimation bias. These large deviations potentially destabilize gradient descent, leading to steady-state misalignment and degraded convergence performance.

This paper introduces the generalized modified Blake-Zisserman (GMBZ) robust loss function [26], and develops generalized modified Blake-Zisserman robust spline adaptive filtering (SAF-GMBZ) algorithm to mitigate these issues. Compared to the SAF-type algorithm, the proposed algorithm achieves superior convergence performance and enhanced robustness in NSI. In the SAF-GMBZ algorithm, the GMBZ exhibits Gaussian-like behavior for inlier points by assigning higher weights to small errors, thus promoting rapid convergence, and reducing search complexity in high-dimensional parameter spaces. Meanwhile, at outlier points, it approximates a uniform distribution, effectively suppressing the influence of abnormal data on the weight update process. This dual property prevents impulsive interference, reduces steady-state misalignment, and improves estimation accuracy.

The partitioning mechanism of GMBZ for inliers and outliers allows the SAF-GMBZ algorithm to dynamically adjust the weight update strategy for spline control-points. This method minimizes the steady-state error while mitigating the outlier-induced weight shift, enhancing the adaptability to nonlinear systems and anomalous data points. Compared with previous SAF algorithms, SAF-GMBZ integrates the nonlinear modeling capability of the spline function with the partition processing advantage of GMBZ, achieving improved convergence performance in GGN environments. Additionally, it demonstrates excellent robustness and stability under impulsive noise conditions, providing a novel and effective solution for adaptive filtering applications.

To ensure the stability of the algorithm, this paper first derives the convergence conditions by analyzing the step-sizes ranges and the steady-state mean-square error (MSE) behavior of SAF-GMBZ. Following this, the proposed algorithm is simulated in NSI and ANC, respectively. Simulation results confirm that SAF-GMBZ outperforms existing SAF methods in GGN environments, while filter-c GMBZ (FcGMBZ) effectively suppresses impulsive noise in ANC application.

The main contributions of this study are summarized as follows:
1) This paper proposes the SAF-GMBZ algorithm, which improves the steady-state accuracy and robustness of the SAF algorithm under GGN.

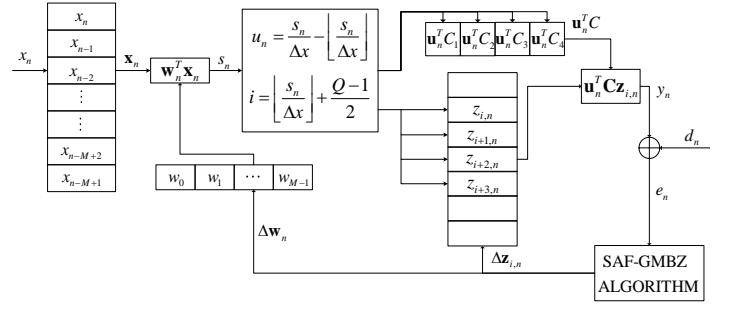

**Fig. 1.** Structure of the SAF-GMBZ algorithm.

2) The mean-square stability analysis of the SAF-GMBZ algorithm is conducted, and convergence step bounds and the steady-state MSE behavior are given.
3) The FcGMBZ algorithm is proposed, experiments in ANC application prove that the FcGMBZ algorithm has accurate convergence and effective noise reduction under impulsive interference environment.

The remainder of this paper is organized as follows: Section II describes the SAF model and the GMBZ criterion. Section III derives the SAF-GMBZ algorithm. Section IV calculates the convergence ranges and steady-state analysis results. Section V presents the simulation results. Finally, Section VI summarizes the paper.

## II. BRIEF REVIEW OF SAF AND GMBZ CRITERION

### A. The Structure of Wiener SAF

Fig. 1 provides a structure of the wiener SAF model, which consists of a linear finite impulse response (FIR) filter cascade with a spline interpolation function. In this model, $n$ represents the instantaneous time and $\mathbf{x}_n = [x_n, x_{n-1}, \ldots, x_{n-N+1}]^T$ represents the input signal of the system with $N$ length, $y_n$ is the output signal of the system, and $s_n$ is the output of the linear network, described as

$$s_n = \mathbf{w}_n^T \mathbf{x}_n \quad (1)$$

where $\mathbf{w}_n = [w_0, w_1, \ldots, w_{N-1}]^T$ denotes the weight vector of the FIR filter.

The local parameter $u_n$ is computed as

$$u_n = \frac{s_n}{\Delta x} - \left\lfloor \frac{s_n}{\Delta x} \right\rfloor \quad (2)$$

where $\Delta x$ represents the horizontal spacing between adjacent control-points within the interpolation interval (with a length of $0 \sim Q$).

The spanning index $i$ is calculated as

$$i = \left\lfloor \frac{s_n}{\Delta x} \right\rfloor + \frac{Q-1}{2} \quad (3)$$

where $\lfloor \cdot \rfloor$ indicates the floor operator, and $Q$ is the total number of control-points.

For the nonlinear function, the input signal $s_n$ is used to compute the parameter $u_n$ and index $i$ through Eq. (2) and (3). The look-up table (LUT) selects the correct control-points vector $\mathbf{z}_i = [z_i, z_{i+1}, z_{i+2}, z_{i+3}]^T$ base on the index $i$. Finally, the output of the SAF $y_n$ is obtained using spline interpolation.

$$y_n = \varphi_i(u_n) = \mathbf{u}_n^T \mathbf{C} \mathbf{z}_{i,n} \quad (4)$$

where $\mathbf{u}_n = [u_n^3, u_n^2, u_n^1, 1]^T$, and $\mathbf{C}$ is the spline base matrix. Among the nonlinear adaptive filters, B-spline [27] and Catmull-Rom (CR) spline [28] are the most widely used. Compared to B-spline, CR-spline exhibits better local approximation properties, providing higher estimation accuracy [4]. Therefore, this paper adopts the CR-spline, expressed as

$$\mathbf{C} = \frac{1}{2}\begin{bmatrix} -1 & 3 & -3 & 1 \\ 2 & -5 & 4 & -1 \\ -1 & 0 & 1 & 0 \\ 0 & 2 & 0 & 0 \end{bmatrix} \quad (5)$$

By taking the derivation of $y_n$ with respect to $u_n$ yields

$$\frac{\partial y_n}{\partial u_n} = \frac{\partial \varphi_i(u_n)}{\partial u_n} = \dot{\mathbf{u}}_n^T \mathbf{C} \mathbf{z}_{i,n} \quad (6)$$

where $\dot{\mathbf{u}}_n = [3u_n^2, 2u_n^1, 1, 0]^T$.

B. GGD and GMBZ Cost Function

This paper primarily evaluates the robustness of the proposed algorithm under GGN environments by introducing the zero-mean generalized gaussian distribution (GGD) as a modeling framework for background noise. The probability density function of the GGD is defined as

$$f(x; \alpha, \sigma^2) = \frac{\alpha}{2\beta \Gamma(1/\alpha)} \exp\left[-\left(\frac{|x|}{\beta}\right)^\alpha\right] \quad (7)$$

where $\beta = \sigma\sqrt{\Gamma(1/\alpha)/\Gamma(3/\alpha)}$ is the scale parameter, $\alpha > 0$ is the shape argument that controls the kurtosis of the distribution, and $\Gamma(\cdot)$ denotes the gamma function. several typical noise distributions of the GGD cover from heavy to light tail, including the Laplacian distribution with $\alpha = 1$, the Gaussian distribution with $\alpha = 2$, the impulse distribution as $\alpha \to 0$.

The GMBZ cost function [26] is defined as

$$J_{GMBZ} = \log(1+\gamma) - E\left\{\log\left[\exp(-\lambda |e_n|^\alpha) + \gamma\right]\right\} \quad (8)$$

where $\gamma > 0$ is the normalization constant, and $\lambda = 1/\beta^\alpha$ is the kernel width parameter [20]. From Eq. (8), when the error is small ($e_n \to 0$), the exponential term approximates $e^{-\lambda e_n^2}$ and resembles a Gaussian-like function, assigning higher weights for small errors and facilitating faster convergence. When $e_n$ is large, the exponential term of the GMBZ is approximated to 0 and the gradient is mainly adjusted by the logarithmic term. Therefore, it exhibits a behavior like a uniform function as the gradient becomes progressively flatter. This prevents extreme data from having an impact on the weight updating process, and improves steady-state accuracy.

Fig. 2 compares several cost functions, including the sign algorithm (SA), the LMS algorithm, the MCC algorithm, the GM-

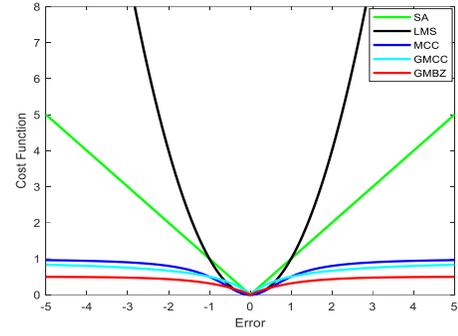

**Fig. 2.** Comparison of cost function with respect to the error.

CC algorithm, and the introduced GMBZ algorithm. While the GMBZ shares exponential characteristics of MCC and GMCC, its additional logarithmic term enhances robustness against heavy-tailed noise. Mathematically, the gradient of GMBZ is given by

$$\frac{\partial J_{GMBZ}(e_n)}{\partial e_n} = -\lambda \alpha e_n^{\alpha-1} \frac{\exp(-\lambda e_n^\alpha)\text{sign}(e_n)}{\exp(-\lambda e_n^\alpha) + \gamma}. \quad (9)$$

The Eq. (9) ensures a smoother gradient decay, making GMBZ highly effective in reducing steady-state errors in non-Gaussian noise environments.

III. PROPOSED ALGORITHM

A. Proposed SAF-GMBZ Algorithm

This section provides the derivation of the SAF-GMBZ algorithm. According to Fig. 1, the *a priori* error is defined as

$$e_n = d_n - y_n = d_n - \varphi_i(u_n). \quad (10)$$

Based on the structure of SAF, the expression of the GMBZ criterion is reformulated as follows

$$J(\mathbf{w}, \mathbf{z}) = \log(1+\gamma) - E\left\{\log\left[\exp(-\lambda |e_n|^\alpha) + \gamma\right]\right\}. \quad (11)$$

According to the gradient descent method and the chain rule, we can obtain

$$\frac{\partial J(\mathbf{w}_n, \mathbf{z}_{i,n})}{\partial \mathbf{w}_n} = -\frac{\lambda \alpha e_n^{\alpha-1} e^{-\lambda e_n^\alpha} \text{sign}(e_n)}{e^{-\lambda e_n^\alpha} + \gamma} \frac{\partial \varphi_i(u_n)}{\partial u_n} \frac{\partial u_n}{\partial s_n} \frac{\partial s_n}{\partial \mathbf{w}_n}$$
$$= -\frac{\lambda \alpha}{\Delta x} \frac{e_n^{\alpha-1} e^{-\lambda e_n^\alpha} \text{sign}(e_n)}{e^{-\lambda e_n^\alpha} + \gamma} \varphi_i'(u_n) \mathbf{x}_n \quad (12)$$

where $\frac{\partial \varphi_i(u_n)}{\partial u_n} = \varphi_i'(u_n) = \dot{\mathbf{u}}_n^T \mathbf{C} \mathbf{z}_{i,n}$, $\frac{\partial s_n}{\partial \mathbf{w}_n} = \mathbf{x}_n$, and $\frac{\partial u_n}{\partial s_n} = \frac{1}{\Delta x}$.

Similarly, the gradient vector at the control-points $\mathbf{z}_{i,n}$ is calculated as

$$\frac{\partial J(\mathbf{w}_n, \mathbf{z}_{i,n})}{\partial \mathbf{z}_{i,n}} = -\frac{\lambda \alpha e_n^{\alpha-1} e^{-\lambda e_n^\alpha} \text{sign}(e_n)}{e^{-\lambda e_n^\alpha} + \gamma} \mathbf{C}^T \mathbf{u}_n \quad (13)$$

where $\frac{\partial \varphi_i(u_n)}{\partial \mathbf{z}_{i,n}} = \mathbf{C}^T \mathbf{u}_n$.

Then the recursions for the weight vector and control-points vector in the developed SAF-GMBZ algorithm are summarized as

$$\mathbf{w}_{n+1} = \mathbf{w}_n + \eta_\mathbf{w} f(e_n) \varphi_i'(u_n) \frac{\mathbf{x}_n}{\Delta x} \quad (14)$$

TABLE I

**Algorithm 1:** Summary of the SAF-GMBZ Algorithm

**Initialization:** $w_0$, $z_0$, $\gamma$, $Q$, $\Delta x$, $N$

1: **for** $n = 0, 1, 2, \ldots$ **do**
2:     Compute the input signal of interpolation function:
$$s_n = \mathbf{w}_n^T \mathbf{x}_n$$
3:     Compute the span index $i$ and the local parameter $u_n$:
$$i = \left\lfloor \frac{s_n}{\Delta x} \right\rfloor + \frac{Q-1}{2}$$
$$u_n = \frac{s_n}{\Delta x} - \left\lfloor \frac{s_n}{\Delta x} \right\rfloor$$
4:     Estimate the output signal of the SAF:
$$y_n = \varphi_i(u_n) = \mathbf{u}_n^T \mathbf{C} \mathbf{z}_{i,n}$$
5:     Compute the error signal:
$$e_n = d_n - y_n$$
6:     Update the weight vector and control points vector:
$$\mathbf{w}_{n+1} = \mathbf{w}_n + \eta_\mathbf{w} f(e_n) \dot{\mathbf{u}}_n^T \mathbf{C} \mathbf{z}_{i,n} \frac{\mathbf{x}_n}{\Delta x}$$
$$\mathbf{z}_{i,n+1} = \mathbf{z}_{i,n} + \eta_\mathbf{z} f(e_n) \mathbf{C}^T \mathbf{u}_n$$
where $f(e_n) = \dfrac{e_n^{\alpha-1} e^{-\lambda e_n^\alpha} \mathrm{sign}(e_n)}{e^{-\lambda e_n^\alpha} + \gamma}$

7: **end for**

$$\mathbf{z}_{i,n+1} = \mathbf{z}_{i,n} + \eta_\mathbf{z} f(e_n) \mathbf{C}^T \mathbf{u}_n \quad (15)$$

With $f(e_n) = \dfrac{e_n^{\alpha-1} e^{-\lambda e_n^\alpha} \mathrm{sign}(e_n)}{e^{-\lambda e_n^\alpha} + \gamma}$, $\eta_\mathbf{w} = \lambda \alpha \mu_\mathbf{w}$ and $\eta_\mathbf{z} = \lambda \alpha \mu_\mathbf{z}$ are the step-sizes parameter.

Finally, Table I presents an overview of the SAF-GMBZ algorithm.

### B. Proposed FcGMBZ Algorithm

Fig. 3 illustrates the schematic diagram of a nonlinear single-channel feed-forward ANC system. In this system, a reference microphone captures the noise signal originating from the noise source and feeds it into the ANC system. The ANC system then processes the reference signal using the proposed FcGMBZ algorithm, generating a control signal with the same amplitude but opposite phase to the noise. This control signal is amplified and played through a loudspeaker, creating a destructive interference pattern that effectively cancels the unwanted noise.

The residual noise captured by the microphone is defined as
$$e_n = d_n - y_n * s_{N,n} \quad (16)$$

where $*$ indicates the convolution operator, and $s_{N,n}$ denotes the impulse response of the secondary channel. In this subsection, the SAF-GMBZ algorithm is extended to the ANC framework, leading to the development of the FcGMBZ algorithm. The weight and control-point updates in the FcGMBZ algorithm are formulated as follows

$$\mathbf{w}_{n+1} = \mathbf{w}_n - \mu_\mathbf{w} \frac{\partial J(\mathbf{w}_n, \mathbf{z}_{i,n})}{\partial \mathbf{w}_n} = \mathbf{w}_n + \eta_\mathbf{w} f(e_n) \mathbf{x}_n' \quad (17)$$

$$\mathbf{z}_{i,n+1} = \mathbf{z}_{i,n} - \mu_\mathbf{z} \frac{\partial J(\mathbf{w}_n, \mathbf{z}_{i,n})}{\partial \mathbf{z}_{i,n}} = \mathbf{z}_{i,n} + \eta_\mathbf{z} f(e_n) \mathbf{C}^T \mathbf{u}_n' \quad (18)$$

where $\mathbf{x}_n' = \dfrac{\varphi_i'(u_n) \mathbf{x}_n}{\Delta x} * s_{N,n}$ and $\mathbf{u}_n' = \mathbf{u}_n * s_{N,n}$.

## IV. PERFORMANCE ANALYSIS

This section analyzes the ranges of the step-sizes and steady-state performance of the algorithm in a system identification model.

### A. Stabilization Range of SAF-GMBZ

In this subsection, the stability range of SAF-GMBZ is derived. Firstly, for the weight vector, the Taylor expansion of the $e_{n+1}$ starting from moment $n$ is described as follows

$$e_{n+1} = e_n + \frac{\partial e_n}{\partial \mathbf{w}_n^T} \Delta \mathbf{w}_n + h.o.t. \quad (19)$$

where $h.o.t.$ denotes the higher-order terms.

By defining $\Delta \mathbf{w}_n = \mathbf{w}_{n+1} - \mathbf{w}_n$, and substituting (14) into the Eq. (19). After simple manipulations, one can yield

$$e_{n+1} = [1 - \eta_\mathbf{w} \lambda \alpha \frac{e_n^{\alpha-2} e^{-\lambda e_n^\alpha} \mathrm{sign}(e_n)}{e^{-\lambda e_n^\alpha} + \gamma} (\varphi_i'^2(u_n) \| \mathbf{x}_n \|^2)] e_n. \quad (20)$$

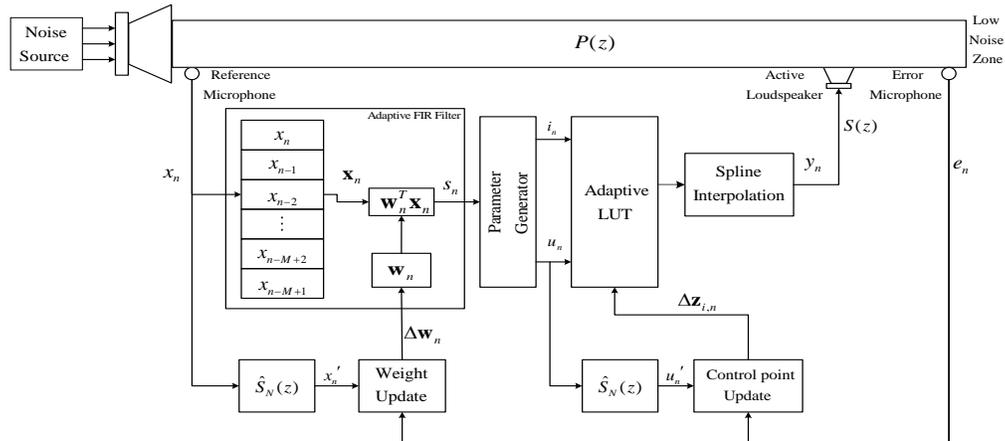

**Fig. 3.** Structure of nonlinear single-channel feed-forward ANC system.

To ensure the algorithm eventually converges, namely the estimation error should be continually decreased throughout the entire iteration process. Therefore, the following condition is satisfied.

$$\left|1-\eta_{\mathbf{w}}\frac{e_n^{\alpha-2}e^{-\lambda e_n^{\alpha}}\text{sign}(e_n)}{e^{-\lambda e_n^{\alpha}}+\gamma}(\varphi_i'^2(u_n)\|\mathbf{x}_n\|^2)\right|\leq 1 \quad (21)$$

which is equivalent to

$$0<\eta_{\mathbf{w}}\leq\frac{2}{\dfrac{e_n^{\alpha-2}e^{-\lambda e_n^{\alpha}}\text{sign}(e_n)}{e^{-\lambda e_n^{\alpha}}+\gamma}\varphi_i'^2(u_n)\|\mathbf{x}_n\|^2}. \quad (22)$$

Similarly, the convergence bounds of the $\eta_{\mathbf{z}}$ can be also described as follows

$$0<\eta_{\mathbf{z}}\leq\frac{2}{\dfrac{e_n^{\alpha-2}e^{-\lambda e_n^{\alpha}}\text{sign}(e_n)}{e^{-\lambda e_n^{\alpha}}+\gamma}\|\mathbf{C}^T\mathbf{u}_n\|^2}. \quad (23)$$

B. Steady-state Performance of SAF-GMBZ

This subsection analyzes the steady-state excess mean-square error (EMSE) performance of the proposed algorithm. The MSE is calculated as the expectation of the squared error:

$$MSE=10\log_{10}\{E[e_n^2]\}. \quad (24)$$

The EMSE is subsequently defined as:

$$\tau=MSE-\sigma_v^2. \quad (25)$$

Fig. 4 illustrates the system identification model used to analyze the SAF algorithm. The system consists of the structure to be identified (denoted by $o$) and the adaptive structure, where $\mathbf{w}_o$ and $\mathbf{z}_o$ indicate the optimal values of the system. When the nonlinear system converges to the steady-state, the following value is derived as $\lim_{n\to\infty}=E\{\mathbf{w}_n\}=\mathbf{w}_o$, and $\lim_{n\to\infty}=E\{\mathbf{z}_{i,n}\}=\mathbf{z}_{i,o}$. This study discusses the case of a linear filter and nonlinear spline function individually. Where $\varepsilon_{\mathbf{w}}$ and $\varepsilon_{\mathbf{z}}$ indicate the a priori error in adaptively adjusting the weights and control-points, respectively

The following assumptions are utilized to simplify the mathematical calculations.

*Assumption 1*: The background noise $v_n$ is stationary with zero-mean and variance $\sigma_v^2$.

*Assumption 2*: The noise $v_n$ is independent of $\mathbf{x}_n$, $s_n$, $\varepsilon_n$, $\varepsilon_{\mathbf{w},n}$ and $\varepsilon_{\mathbf{z},n}$.

*Assumption 3*: At steady-state, the update term $f(e_n)$ and a priori error $\varepsilon_n$ are independent of $\varphi_i'(u_n)$, $\varphi_i'^2(u_n)$, $\|\mathbf{x}_n\|^2$ and $\|\mathbf{C}^T\mathbf{u}_n\|^2$.

Next, by defining the weight error vector as $\mathbf{v}_n^{(\mathbf{w})}=\mathbf{w}_o-\mathbf{w}_n$ and the control-points error vector as $\mathbf{v}_n^{(\mathbf{z})}=\mathbf{z}_{i,o}-\mathbf{z}_{i,n}$, and referring to Fig. 3, the $\varepsilon_n$ is computed by

$$\begin{aligned}\varepsilon_n&=d_n-y_n=\varphi_o(t_n)-\varphi(s_n)=\varphi_i(u_{t,n})-\varphi_i(u_{s,n})\\&=\mathbf{u}_{t,n}^T\mathbf{C}\mathbf{z}_{i,o}-\mathbf{u}_{s,n}^T\mathbf{C}\mathbf{z}_{i,n}=\varepsilon_{\mathbf{w},n}+\varepsilon_{\mathbf{z},n}\end{aligned} \quad (26)$$

where $\mathbf{u}_{t,n}=[u_{t,n}^3,u_{t,n}^2,u_{t,n}^1,1]^T$, $\mathbf{u}_{s,n}=[u_{s,n}^3,u_{s,n}^2,u_{s,n}^1,1]^T$, with $u_{t,n}$ and $u_{s,n}$ indicating the local parameter of $t_n$ and $s_n$, respectively. From (2) we can achieve that

$$\begin{aligned}u_{t,n}-u_{s,n}&=\frac{t_n}{\Delta x}-\left\lfloor\frac{t_n}{\Delta x}\right\rfloor-\frac{s_n}{\Delta x}+\left\lfloor\frac{s_n}{\Delta x}\right\rfloor=\frac{t_n}{\Delta x}-\frac{s_n}{\Delta x}\\&\approx\frac{(\mathbf{w}_o-\mathbf{w}_n)^T\mathbf{x}_n}{\Delta x}=\frac{v_n^{(\mathbf{w})T}\mathbf{x}_n}{\Delta x}.\end{aligned} \quad (27)$$

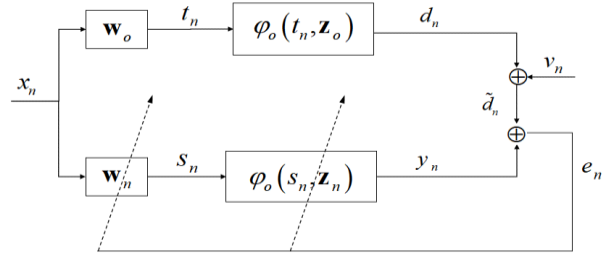

**Fig. 4.** Block diagram of system identification.

When only the weight error of the linear filter is adaptively adjusted, the following results are derived as

$$\begin{aligned}\varepsilon_{\mathbf{w},n}&=(\mathbf{u}_{t,n}-\mathbf{u}_{s,n})^T\mathbf{C}\mathbf{z}_{i,o}=\mathbf{u}_{t,n}^T\mathbf{C}\mathbf{z}_{i,o}-\mathbf{u}_{s,n}^T\mathbf{C}\mathbf{z}_{i,o}\\&\approx u_{t,n}\mathbf{c}_3\mathbf{z}_{i,o}+\mathbf{c}_4\mathbf{z}_{i,o}-u_{s,n}\mathbf{c}_3\mathbf{z}_{i,o}-\mathbf{c}_4\mathbf{z}_{i,o}\\&=(u_{t,n}-u_{s,n})\mathbf{c}_3\mathbf{z}_{i,o}=\frac{\mathbf{c}_3\mathbf{z}_{i,o}}{\Delta x}\mathbf{v}_n^{(\mathbf{w})T}\mathbf{x}_n.\end{aligned} \quad (28)$$

Using the definition of $\mathbf{v}_n^{(\mathbf{w})}$, we subtract both sides of (14) from $\mathbf{w}_o$ and get

$$\mathbf{v}_{n+1}^{(\mathbf{w})}=\mathbf{v}_n^{(\mathbf{w})}-\eta_{\mathbf{w}}f(e_n)\frac{\varphi_i'(u_n)}{\Delta x}\mathbf{x}_n. \quad (29)$$

Now evaluating the energies of two sides of (29), and taking the expectation yields

$$\begin{aligned}E\{\|\mathbf{v}_{n+1}^{(\mathbf{w})}\|^2\}&=E\{\|\mathbf{v}_n^{(\mathbf{w})}\|^2\}-2\eta_{\mathbf{w}}E\left\{\frac{\varphi_i'(u_n)}{\mathbf{c}_3\mathbf{z}_{i,n}}f(e_n)\varepsilon_{\mathbf{w},n}\right\}\\&+\eta_{\mathbf{w}}^2E\left\{\frac{\varphi_i'^2(u_n)}{\Delta x^2}\|\mathbf{x}_n\|^2f^2(e_n)\right\}.\end{aligned} \quad (30)$$

When the algorithm converges to the steady-state, the condition $\lim_{n\to\infty}E\{\|\mathbf{v}_{n+1}^{(\mathbf{w})}\|^2\}=\lim_{n\to\infty}E\{\|\mathbf{v}_n^{(\mathbf{w})}\|^2\}$ is established, and the following result is obtained

$$2\Delta x^2E\left\{\frac{\varphi_i'(u_n)}{\mathbf{c}_3\mathbf{z}_{i,n}}\right\}E\{f(e_n)\varepsilon_{\mathbf{w},\infty}\}=\eta_{\mathbf{w}}E\{\varphi_i'^2(u_n)\|\mathbf{x}_n\|^2\}E\{f^2(e_n)\}. \quad (31)$$

By carrying the Taylor series expansion of $E\{f(e_n)\varepsilon_{\mathbf{w},\infty}\}$ and $E\{f^2(e_n)\}$ with respect to $v_n$ respectively and employing *Assumption 1*, leads to

$$E\{f(e_n)\varepsilon_{\mathbf{w},n}\} = E\{f(\varepsilon_{\mathbf{w},n}+v_n)\varepsilon_{\mathbf{w},n}\}$$
$$\approx E\{f(v_n)\varepsilon_{\mathbf{w},n}+f'(v_n)\varepsilon_{\mathbf{w},n}^2\} \approx E\{f'(v_n)\}E\{\varepsilon_{\mathbf{w},n}^2\} \quad (32)$$

and

$$E\{f^2(e_n)\} = E\{f^2(\varepsilon_{\mathbf{w},n}+v_n)\} \approx E\{f^2(v_n)\}+$$
$$E\{|f'(v_n)|^2\}E\{\varepsilon_{\mathbf{w},n}^2\}+E\{f(v_n)\}E\{f''(v_n)\}E\{\varepsilon_{\mathbf{w},n}^2\}. \quad (33)$$

Substituting (32) and (33) into (31) yields (34) (at the bottom of the page).

Following a similar method for the calculation of (27), we have

$$\varepsilon_{\mathbf{z},n} = \mathbf{u}_{s,n}^T \mathbf{C}(\mathbf{z}_{i,0}-\mathbf{z}_{i,n}) = \mathbf{u}_{s,n}^T \mathbf{C}\mathbf{v}_n^{(\mathbf{z})} = \mathbf{v}_n^{(\mathbf{z})T}\mathbf{C}^T\mathbf{u}_{s,n}. \quad (35)$$

Using the definition of $\mathbf{v}_n^{(\mathbf{z})}$, subtracting both sides of (15) from $\mathbf{z}_o$ and get

$$\mathbf{v}_{n+1}^{(\mathbf{z})} = \mathbf{v}_n^{(\mathbf{z})} - \eta_{\mathbf{z}}f(e_n)\mathbf{C}^T\mathbf{u}_n. \quad (36)$$

When the algorithm arrives at the steady-state, the condition $\lim_{n\to\infty}E\{\|\mathbf{v}_{n+1}^{(\mathbf{z})}\|^2\} = \lim_{n\to\infty}E\{\|\mathbf{v}_n^{(\mathbf{z})}\|^2\}$ is established, and the following result is given by

$$2E\{f(e_n)\varepsilon_{\mathbf{z},\infty}\} = \eta_{\mathbf{z}}E\{\|\mathbf{C}^T\mathbf{u}_n\|^2\}E\{f^2(e_n)\}. \quad (37)$$

Substituting the Taylor expansion results in (32) and (33) into (37), the result is shown in (38) (at the bottom of the page).

Finally, when the algorithm converges to steady-state $n\to\infty$, there exists $u_{t,n}\approx u_{s,n}$ and $\mathbf{z}_{i,0}\approx\mathbf{z}_{i,n}$. From (28) and (35), we observed that the $E\{\varepsilon_{\mathbf{w},\infty}\varepsilon_{\mathbf{z},\infty}\}$ can be omitted to give the following EMSE results

$$\tau = E\{\varepsilon_\infty^2\} = E\{(\varepsilon_{\mathbf{w},\infty}+\varepsilon_{\mathbf{z},\infty})^2\}$$
$$= E\{\varepsilon_{\mathbf{w},\infty}^2\} + E\{\varepsilon_{\mathbf{z},\infty}^2\} + 2E\{\varepsilon_{\mathbf{w},\infty}\varepsilon_{\mathbf{z},\infty}\} \quad (39)$$
$$= E\{\varepsilon_{\mathbf{w},\infty}^2\} + E\{\varepsilon_{\mathbf{z},\infty}^2\}.$$

## V. SIMULATION RESULTS

This section presents the simulation results for the proposed algorithm. The effectiveness of the proposed methods is assessed by comparing them with existing algorithms under various noise environments. The unknown system utilized for NSI is a wiener system, consisting of a linear component $\mathbf{w}_0 = [0.6, -0.4, 0.25, -0.15, 0.1, -0.05, 0.001]^T$, and a nonlinear interpolation function realized by a 23-point length LUT $\mathbf{z}_0$. The sampling interval $\Delta x$ is set to 0.2, and $\mathbf{z}_0 = [-2.20, -2.00, -1.80, -1.60, -1.40, -1.20, -1.00, -0.80, -0.91, -0.40, 0.20, 0.00, 0.15, 0.58, 1.00, 1.00, 1.20, 1.40, 1.60, 1.80, 2.00, 2.20]$. The input signal $\mathbf{x}_n$ is generated by filtering a white Gaussian signal with unit variance through a colored signal filter

$$H(z) = \frac{\sqrt{1-\zeta^2}}{1-\zeta z^{-1}} \quad (40)$$

where $0\leq\zeta<1$ represents the similarity between two adjacent input samples, with $\zeta=0.1$.

$$E\{\varepsilon_{\mathbf{w},n}^2\} = \frac{\eta_{\mathbf{w}}E\{\varphi_i'^2(u_n)\|\mathbf{x}_n\|^2\}E[f^2(v_n)]}{2\Delta x^2 E\left\{\frac{\varphi_i'(u_n)}{\mathbf{c}_3\mathbf{z}_{i,n}}\right\}E[f'(v_n)] - \eta_{\mathbf{w}}E\{\varphi_i'^2(u_n)\|\mathbf{x}_n\|^2\}E\left[f(v_n)f''(v_n)+|f'(v_n)|^2\right]} \quad (34)$$

$$E\{\varepsilon_{\mathbf{z},n}^2\} = \frac{\eta_{\mathbf{z}}E\{\|\mathbf{C}^T\mathbf{u}_n\|^2\}E[f^2(v_n)]}{2E[f'(v_n)] - \eta_{\mathbf{z}}E\{\|\mathbf{C}^T\mathbf{u}_n\|^2\}E\left[f(v_n)f''(v_n)+|f'(v_n)|^2\right]}. \quad (38)$$

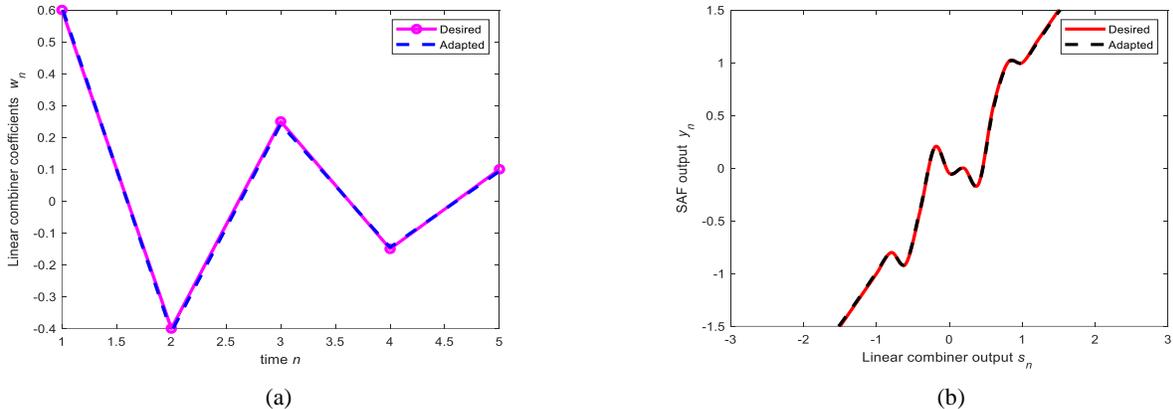

(a)      (b)

**Fig. 5.** The identification results of the classical nonlinear system using SAF-GMBZ (SNR=30dB, $\zeta=0.1$): (a) Comparison of the desired and adapted linear combiner coefficient $\mathbf{w}_n$, and (b) Comparison of the desired and adapted nonlinearity profile.

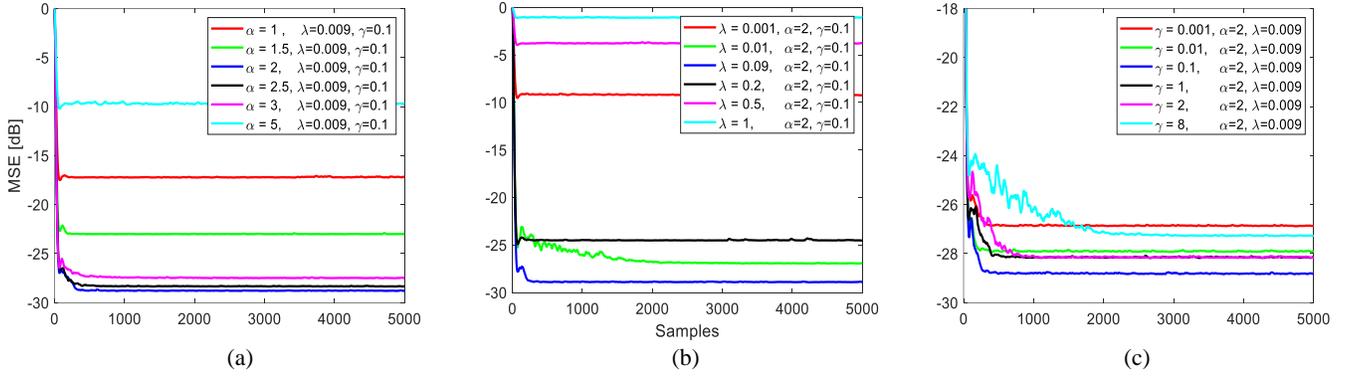

**Fig. 6.** MSE curves of the proposed algorithm with different parameters: (a) $\alpha$, (b) $\lambda$, and (c) $\gamma$.

TABLE II
Simulation Parameters for NSI

|  | Gaussian |  | Laplace |  | Uniform |  | Binary |  | Generalized Gaussian |  |
|---|---|---|---|---|---|---|---|---|---|---|
| Algorithm | $\mu_w$ | $\mu_z$ | $\mu_w$ | $\mu_z$ | $\mu_w$ | $\mu_z$ | $\mu_w$ | $\mu_z$ | $\mu_w$ | $\mu_z$ |
| SAF-LMS | 0.04 | 0.08 | 0.05 | 0.08 | 0.05 | 0.05 | 0.009 | 0.25 | 0.09 | 0.5 |
| SAF-MCC ($\sigma=1.5$) | 0.03 | 0.06 | 0.04 | 0.09 | 0.03 | 0.08 | 0.01 | 0.08 | 0.005 | 0.02 |
| SAF-GMCC ($\alpha=2$, $\beta=3$) | 0.005 | 0.04 | 0.003 | 0.007 | 0.0009 | 0.03 | 0.003 | 0.03 | 0.03 | 0.09 |
| SAF-GMBZ ($\alpha=2$, $\lambda=0.09$, $\gamma=0.1$) | 0.08 | 0.9 | 0.1 | 0.8 | 0.02 | 0.05 | 0.04 | 0.5 | 0.04 | 0.08 |

In this study, we consider five different types of noise distributions, all of which follow the GGD, at a fixed signal-to-noise ratio (SNR) of 30dB. Among them, four special cases are considered: 1) Gaussian distribution: the probability density function follows a normal distribution with zero-mean and $\sigma_v^2 = 0.001$. 2) Binary distribution: taking values of $\{-1,1\}$, with probabilities P(-1)=0.65 and P(1)=0.35. 3) Laplace distribution: with zero-mean and $\sigma_v^2 = 0.001$, compared with the Gaussian distribution, it has sharper peaks and heavier tails. 4) Uniform distribution: distributed in the range [-0.1,0.8] with constant probability density. In addition, we further consider a general GGN case by setting the shape parameter to an arbitrary value, in order to verify the algorithm's adaptability across varying statistical characteristics.

To ensure a fair comparison, the step-sizes parameters for each algorithm are carefully selected to ensure comparable initial convergence rates, allowing an objective evaluation of steady-state performance. These parameters are summarized in Tables II and III. Each experiment is averaged over 100 independent trials to ensure statistical reliability.

A. NSI Application

The experiments in NSI focus on the comparison of the learning performance of the SAF-GMBZ algorithms in system identification models. Fig. 5 illustrates the identification results of the SAF-GMBZ algorithm for a nonlinear system. Specifically, Fig. 5 (a) illustrates the evolution of the linear filter coefficients $\mathbf{w}_n$ in the SAF-GMBZ adaptive model compared to the desired model. The close match between the two curves confirms the accurate parameter estimation capability of the SAF-GMBZ algorithm. Fig. 5 (b) presents a comparison of the nonlinear mapping functions, validating that the SAF-GMBZ algorithm effectively captures the system's nonlinear characteristics. The high consistency between the two curves further validates the effectiveness of the SAF-GMBZ algorithm in NSI.

*1) Effect of Parameters:* This subsection investigates the impact of GMBZ robust parameters $\alpha$, $\lambda$, and $\gamma$ on the performance of the proposed algorithm. A series of experiments was conducted under fixed and empirically tuned step-sizes settings ($\mu_w = 0.08$, $\mu_z = 0.9$). Each parameter was varied independently while keeping the others constant, and the resulting MSE trends were recorded, as illustrated in Fig. 6.

As shown in Fig. 6 (a), the parameter $\alpha$ significantly influences the steady-state performance of the SAF-GMBZ algorithm. A smaller value of $\alpha$ leads to slower convergence and higher steady-state error. As $\alpha$ increases, both convergence speed and steady-state MSE improve, reaching optimal performance around $\alpha = 2$. Fig. 6 (b) reveals that $\lambda$ has a pronounced effect on steady-state error. When $\lambda = 0.09$, the algorithm achieves a favorable trade-off between convergence speed and steady-state accuracy. According to Fig. 6 (c), the algorithm reaches the optimal steady-state performance when $\gamma = 0.1$. Furthermore, even under extremely small or large of $\gamma$, the algorithm maintains low steady-state error, indicating strong robustness and adaptability. Based on these observations, the optimal parameters $\alpha = 2$, $\lambda = 0.09$, and $\gamma = 0.1$ are adopted in Fig. 7.

*2) Performance Comparison of Different Algorithms:* Fig. 7 compares the MSE results of SAF-GMBZ with comparative algorithms under different noise conditions. For the step-sizes presented in Table II, we ensured that all algorithms exhibit as the steady-state error. As illustrated in Fig. 7, SAF-GMBZ consistently exhibits superior convergence performance across all noise conditions, achieving faster convergence and higher steady-state accuracy than comparative algorithms. These results align with the analysis in Fig. 2, demonstrating that the GMBZ function enhances robustness and effectively mitigates convergence instability caused by abnormal noise. This observation confirms the superiority of the SAF-GMBZ algorithm.

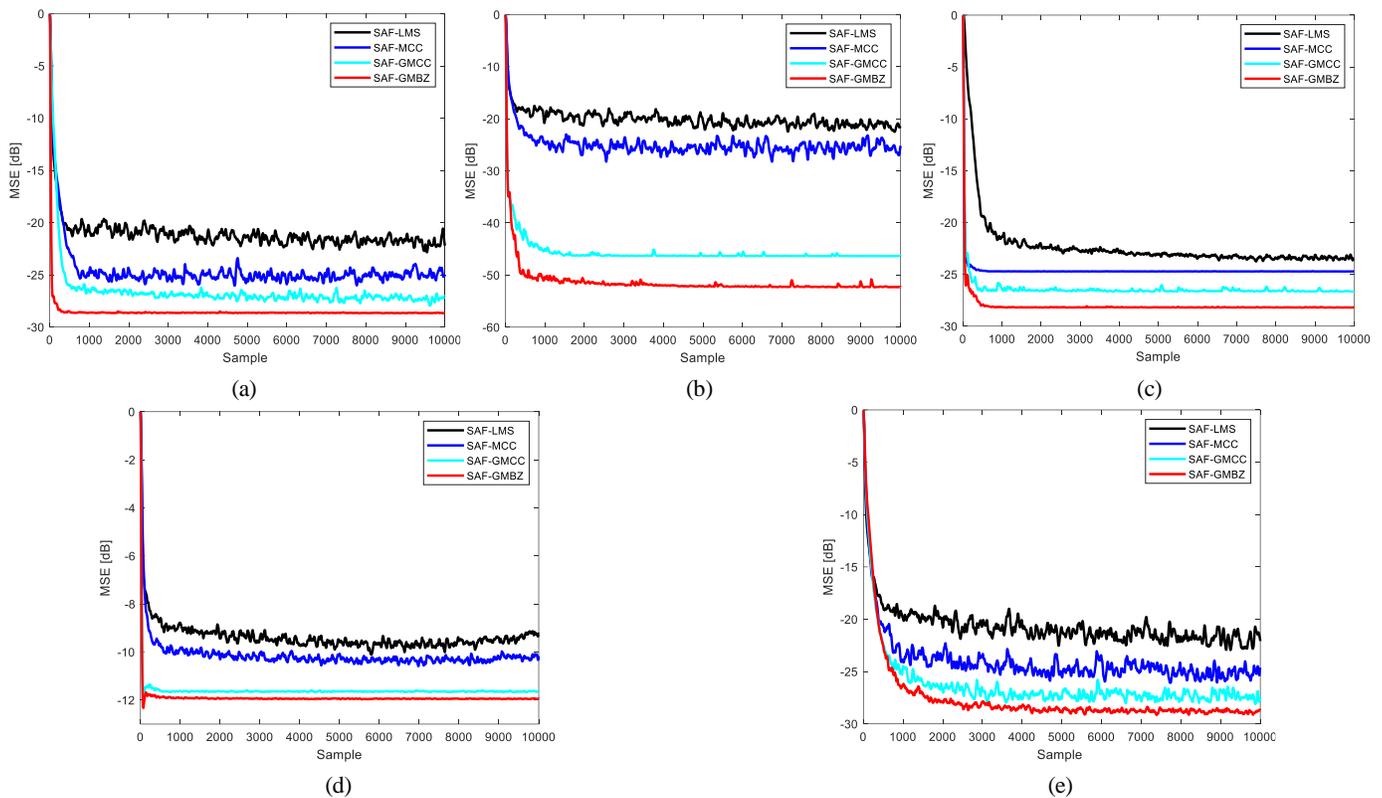

**Fig. 7.** MSE curves of the competitive algorithms under different background noises: (a) Gaussian, (b) Binary, (c) Laplace, (d) Uniform, and (e) Generalized Gaussian ($\alpha = 0.3$).

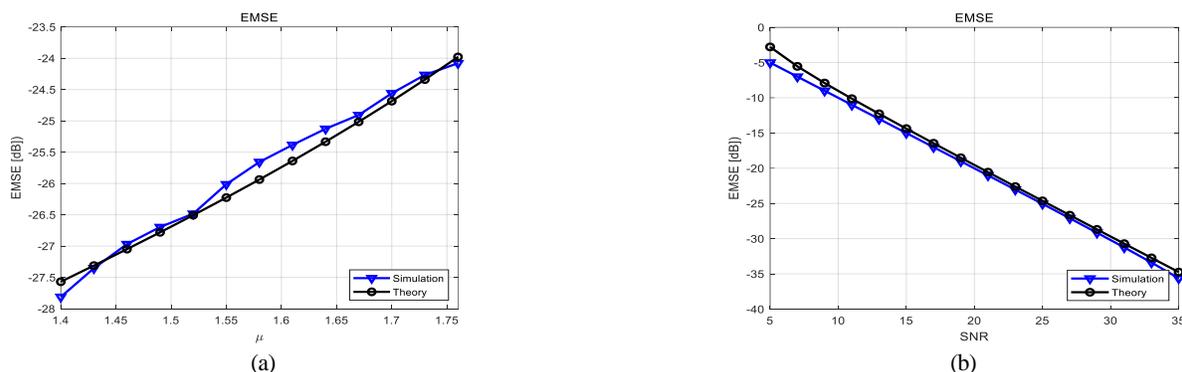

**Fig. 8.** Comparison of theoretical and simulated values of the proposed algorithm: (a) different step-sizes, and (b) different SNRs.

### B. Theoretical Validation

To verify the correctness of the theoretical analysis, Fig. 8 investigates the theoretical and simulated steady-state EMSE curves under various step-sizes and SNR levels. In Fig. 8 (a), the EMSE values gradually increase with the step-sizes, which align with the analysis in (34) and (38). Specifically, as the step-sizes increase, the algorithm's magnitude increases, leading to a higher steady-state error. This occurs because larger step-sizes introduce greater estimation bias, negatively impacting convergence accuracy. Fig. 8 (b) illustrates a decreasing trend in steady-state error as the SNR increases, as higher SNR levels reduce the impact of noise. The simulation results exhibit a strong agreement between the theoretical and simulated curves across various SNR levels and step-sizes, further validating the accuracy of the analysis. Minor deviations between theoretical and simulated EMSE may arise due to the independence and linearization approximations in Assumption 3. In addition, the discrete spline implementation and use of finite data or Monte Carlo trials may slightly affect the steady-state EMSE, without altering the overall agreement.

### C. ANC Application

This subsection tests the noise reduction effect of a single-channel ANC application. We compare the proposed approach with other existing algorithms and utilize the average noise reduction (ANR) metric as a criterion for evaluating the performance of the algorithm, which is given by the following expression

TABLE III
Simulation Parameters for ANC

| Algorithm | $\alpha=2$ | | $\alpha=1.8$ | | $\alpha=1.7$ | | $\alpha=1.5$ | |
|---|---|---|---|---|---|---|---|---|
| | $\mu_w$ | $\mu_z$ | $\mu_w$ | $\mu_z$ | $\mu_w$ | $\mu_z$ | $\mu_w$ | $\mu_z$ |
| SAF-LMS | 0.3 | 0.00001 | 0.3 | 0.00001 | 0.3 | 0.00001 | 0.3 | 0.00001 |
| SAF-SNLMS | 0.3 | 0.00001 | 0.3 | 0.00001 | 0.3 | 0.00001 | 0.3 | 0.00001 |
| SAF-MCC ($\sigma=1.5$) | 0.2 | 0.00008 | 0.2 | 0.00008 | 0.2 | 0.0001 | 0.2 | 0.0001 |
| SAF-GMC ($\sigma=2.1$, $p=2.2$) | 0.1 | 0.00001 | 0.3 | 0.00001 | 0.3 | 0.00001 | 0.3 | 0.00001 |
| SAF-GMCC ($\alpha=2$, $\beta=3$) | 0.1 | 0.00001 | 0.3 | 0.00001 | 0.3 | 0.0001 | 0.3 | 0.00001 |
| SAF-GMBZ ($\alpha=2$, $\lambda=6$, $\gamma=0.1$) | 0.1 | 0.00001 | 0.3 | 0.00001 | 0.1 | 0.0001 | 0.2 | 0.0001 |

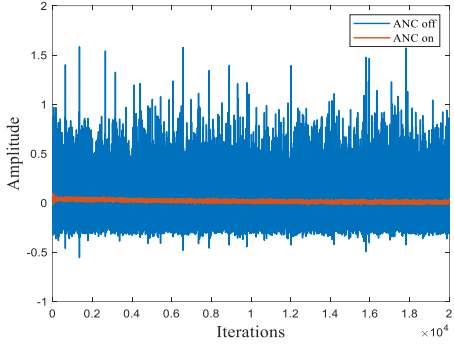

**Fig. 9.** The reference noise and residual noise waveforms of the proposed robust ANC algorithm.

$$\text{ANR}_n = 20\log(A_{e,n}/A_{d,n}) \quad (41)$$
$$A_{e_n} = \chi A_{e_{n-1}} + (1-\chi)|e_n|, A_{e,0} = 0 \quad (42)$$
$$A_{d_n} = \chi A_{d_{n-1}} + (1-\chi)|d_n|, A_{d,0} = 0 \quad (43)$$

where $\chi$ is the forgetting factor, with $\chi \to 1$, $\chi \neq 1$, $\chi=0.999$. Assume that the SNR of the measured noise is 30dB, and perform 100 independent trials. In this experiment, the noise source is generated by the $\alpha$-stable distribution [29]. The α-stable distribution is a heavy-tailed distribution, typically used to model impulsive noise, defined as

$$\varphi(t) = \exp\left(j\vartheta t - |\varepsilon t|^\alpha (1+j\beta\text{sign}(t)\tan\pi\alpha/2)\right) \quad (44)$$

where $0 < \alpha \leq 2$ is the characteristic exponent that determines the noise intensity. The noise captured by the error microphone is given by the following formula

$$d_n = u_{n-2} + \kappa u_{n-2}^2 - \delta u_{n-1}^3 \quad (45)$$

where $\kappa=0.08$, $\delta=0.04$, $u_n = x_n * p_n$, and $p_n$ denotes the impulse response of the primary channel with transfer function. The transfer functions with non-minimum phase primary and secondary channels are

$$P(z) = z^{-3} - 0.3z^{-4} + 0.2z^{-5} \quad (46)$$
$$S(z) = z^{-2} + 1.5z^{-3} - z^{-4} \quad (47)$$

The criteria for selecting the step-sizes parameters for the competitive algorithms in this subsection are consistent with those in Table II, with the specific parameters provided in Table III. Fig. 9 illustrates the waveforms of the reference noise and the residual noise under impulsive interference (with $\alpha=2$). The FcGMBZ method effectively suppresses noise, maintaining low residual noise levels, and demonstrates strong robustness against impulsive disturbances.

Comparison of the ANR performance of different ANC algorithms under various impulsive noise levels are shown in Fig. 10. In Fig. 10 (a), the FcSNLMS, FcMCC, FcGMC, FcGMCC, and FcGMBZ algorithms exhibit effective results in noise control. Among them, the FcGMBZ algorithm achieves the highest convergence accuracy, has the best noise reduction effect, and exhibits stronger robustness. To further verify the advancement of the proposed algorithms, experiments were performed with varying $\alpha$ values (set $\alpha=1.8, 1.7, 1.5$), as shown in Figs. 10 (b), (c), and (d). As the intensity of impulse noise increases, conventional methods struggle to maintain performance, whereas the FcGMBZ consistently achieves the lowest residual noise levels, confirming its superior noise reduction capability.

## VI. CONCLUSION

This paper introduced the SAF-GMBZ algorithm, which enhances the robustness and steady-state accuracy of spline adaptive filtering in GGN environments. To ensure the algorithm's stability, we derived the convergence step-sizes bound and conducted a theoretical analysis of its steady-state EMSE performance. Furthermore, the FcGMBZ algorithm is proposed to improve noise reduction in ANC application, particularly under impulsive noise environment. Simulation results demonstrated that the SAF-GMBZ algorithm achieves superior learning performance and enhanced robustness in GGN environment. Additionally, the FcGMBZ algorithm consistently outperforms existing methods, effectively mitigating impulsive noise, and achieving high steady-state accuracy under impulsive noise. The proposed SAF-GMBZ and FcGMBZ algorithm provide efficient and robust solutions for adaptive filtering in non-Gaussian noise, highlighting its excellent anti-interference capability and stability.

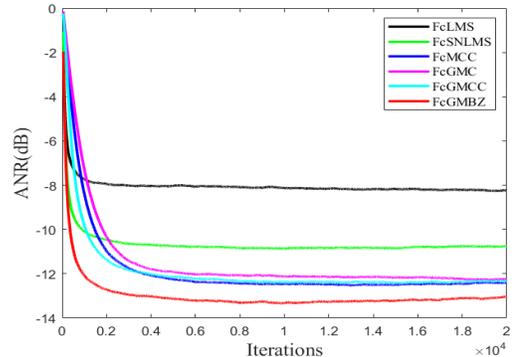

(a)

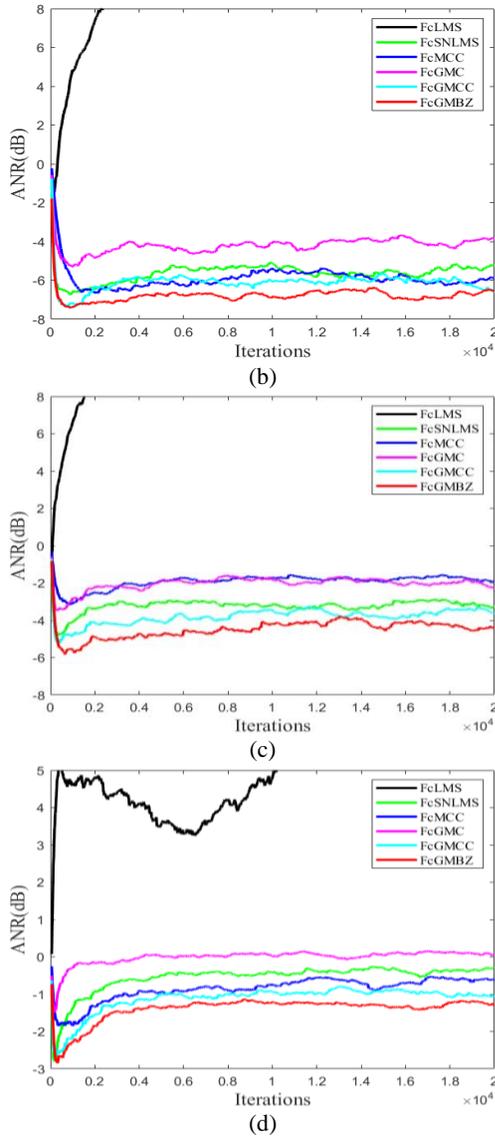

**Fig. 10.** Comparison of ANR under different $\alpha$ impulsive noises ($\beta=0$, $\varepsilon=1$, $\vartheta=0$): (a) $\alpha=2$, (b) $\alpha=1.8$, (c) $\alpha=1.7$, and (d) $\alpha=1.5$.

REFERENCES

[1] H. Zhao and B. Chen, Efficient Nonlinear Adaptive Filters: Design, Analysis and Applications. *Cham, Switzerland: Springer*, 2023.
[2] S. S. Haykin, Adaptive Filter Theory. *Noida, India: Pearson Educ.*, 2008.
[3] Y. G. Dzhugashvili and M. A. Svinin, "Adaptive affine projection algorithm for adaptive filtering," *IEEE Trans. Signal Process.*, vol. 55, no. 2, pp. 550–554, 2007.
[4] M. Scarpiniti, D. Comminiello, R. Parisi and A. Uncini, "Nonlinear spline adaptive filtering," *Signal Process*, vol. 93, no. 4, pp. 772–783, 2013.
[5] J. M. Le Caillec, "Spectral inversion of second order volterra models based on the blind identification of wiener models," *Signal Process.*, vol. 91, no. 11, pp. 2541–2555, 2011.
[6] J. Gong and B. Yao, "Neural network adaptive robust control of nonlinear systems in semistrict feedback form," *Automatica.*, vol. 37, no. 8, pp. 1149–1160, 2001.
[7] W. Guo and Y. Zhi, "Nonlinear spline versoria prioritization optimization adaptive filter for alpha-stable clutter," *IEEE Trans. Aerosp. Electron. Syst.*, vol. 59, no. 1, pp. 734–744, 2023.
[8] Y. Gao, H. Zhao, R. Zhu, "Least mean p-power Hammerstein spline adaptive filtering algorithm: formulation and analysis," *IEEE Trans. Aerosp. Electron. Syst.,* vol. 60, no. 5, pp. 6275–6283, Oct. 2024.
[9] M. Scarpiniti, D. Comminiello, R. Parisi, and A. Uncini, "Nonlinear system identification using IIR spline adaptive filters," *Signal Process.*, vol. 108, pp. 30–35, Mar. 2015.
[10] M. Scarpiniti, D. Comminiello, R. Parisi and A. Uncini, "Hammerstein uniform cubic spline adaptive filters: Learning and convergence properties," *Signal Process.*, vol. 100, pp. 112–123, 2014.
[11] M. Scarpiniti, D. Comminiello, R. Parisi, and A. Uncini, "Novel cascade spline architectures for the identification of non-linear systems," *IEEE Trans. Circuits Syst. I*, vol. 62, no. 7, pp. 1825–1835, Jul. 2015.
[12] M. Scarpiniti, D. Comminiello, G. Scarano, et al., "Steady-state performance of spline adaptive filters," *IEEE Trans. Signal Process.*, vol. 64, no. 4, pp. 816–828, 2015.
[13] S. Guan and Z. Li, "Normalized spline adaptive filtering algorithm for nonlinear system identification," *Neural Process. Lett.*, vol. 46, no. 2, pp. 595–607, 2017.
[14] C. Liu, Z. Zhang, and X. Tang, "Sign normalized Hammerstein spline adaptive filtering algorithm in an impulsive noise environment," *Circuits Syst. Signal Process.*, vol. 38, no. 2, pp. 891–903, 2019.
[15] C. Liu, Z. Zhang, and X. Tang, "Sign-normalized IIR spline adaptive filtering algorithms for impulsive noise environments," *Neural Process. Lett.*, vol. 50, no. 1, pp. 477–496, 2019.
[16] M. Rathod, V. Patel and N. George, "Generalized spline nonlinear adaptive filters," *Expert Syst. Appl.*, vol. 83, pp. 122–130, 2017.
[17] L. Yang, J. Liu, Z. Zhao, R. Yan and X. Chen, "Interval variable step-size spline adaptive filter for the identification of nonlinear block-oriented system," *Nonlinear Dyn.*, vol. 98, no. 3, pp. 1629–1643, 2019.
[18] A. Saenmuang and S. Sitjongsataporn, "Spline adaptive filtering based on variable leaky least mean square algorithm," *in Proc. 2020 8th Int. Electr. Eng. Congr. (iEECON),* Chiang Mai, Thailand, 2020, pp. 1–4.
[19] Y. Gao, H. Zhao, Y. Zhu, et al., "The q-gradient LMS spline adaptive filtering algorithm and its variable step-size variant," *Information Sciences.*, vol. 658, pp. 521–534, 2023.
[20] A. Singh and J. C. Principe, "Using correntropy as a cost function in linear adaptive filters," *in Proc. 2009 Int. Joint Conf. Neural Networks*, Atlanta, GA, USA, 2009, pp. 2950–2955.
[21] S. Peng, Z. Wu, X. Zhang, and B. Chen, "Nonlinear spline adaptive filtering under maximum correntropy criterion," *in TENCON 2015-2015 IEEE Region 10 Conference,* Macao, China, 2015, pp. 1–5.
[22] W. Wang, H. Zhao, X. Zeng, et al., "Steady-state performance analysis of nonlinear spline adaptive filter under maximum correntropy criterion," *IEEE Trans. Circuits Syst. II: Express Briefs.*, vol. 67, no. 6, 1154–1158, 2019.
[23] C. Liu, C. Peng, X. Tang, et al. "Two variants of the IIR spline adaptive filter for combating impulsive noise," *EURASIP J. Adv. Signal Process.*, vol. 2019, no. 8, pp. 1–11, 2019.
[24] W. Guo and Y. Zhi, "Nonlinear spline adaptive filtering against non-gaussian noise," *Circuits Syst. Signal Process.*, vol. 41, pp. 579–596, 2022.
[25] Y. Gao, H. Zhao, Y. Zhu, et al., "Spline adaptive filtering algorithm-based generalized maximum correntropy and its application to nonlinear active noise control," *Circuits Syst. Signal Process.*, vol. 42, pp. 6636–6659, 2023.
[26] M. Kumar, M. L. N. S. Karthik and N. V. George, "Generalized modified Blake–Zisserman robust sparse adaptive filters," *IEEE Trans. Syst., Man, Cybern., Syst.*, vol. 53, no. 1, pp. 647–652, 2023.
[27] H. Prautzsch, B. Wolfgang, and M. Paluszny, Bézier and B-Spline Techniques. Berlin, Germany: Springer-Verlag, 2002.
[28] E. Catmull, R. Rom, Ch. "A class of local interpolating splines," *Comput. Aided Geom. Des.*, 1974, pp. 317–326.
[29] S. Talebi, S. Godsill, and D. Mandic, "Filtering structures for α-stable systems," *IEEE Control Syst. Lett.*, vol. 7, pp. 553–558, 2023.

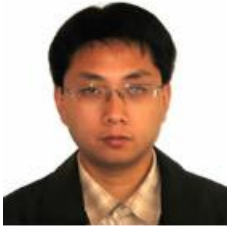

**Haiquan Zhao** He received a B.S. degree in applied mathematics and an M.S. and Ph.D. in signal and information processing from Southwest Jiaotong University, Chengdu, China, in 1998, 2005, and 2011, respectively. Since 2012, he has been a Professor at the School of Electrical Engineering, Southwest Jiaotong University. From 2015 to 2016, he worked as a visiting scholar at the University of Florida, Gainesville, FL, USA. He is the author or co-author of more than 280 international journal papers (SCI indexed) and owns 70 invention patents. His current research interests include information theoretical learning, adaptive filters, adaptive networks, active noise control, Kalman filters, machine learning, and artificial intelligence. Prof. Zhao has won several provincial and ministerial awards and many best paper awards at international conferences or IEEE TRANSACTIONS. He has served as an Active Reviewer for several IEEE TRANSACTIONS, IET series, signal processing, and other international journals. He is currently a Handling Editor of Signal Processing, and also an Associate Editor of IEEE TRANSACTIONS ON AUDIO, SPEECH, AND LANGUAGE PROCESSING, IEEE TRANSACTIONS ON SYSTEMS, MAN AND CYBERNETICS: SYSTEM, IEEE SIGNAL PROCESSING LETTERS, IEEE SENSORS JOURNAL, and IEEE OPEN JOURNAL OF SIGNAL PROCESSING.

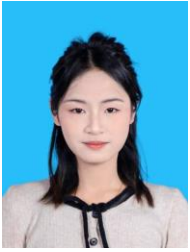

**Bei Xu** received the B.E. degree in automation from Southwest Minzu University, Chengdu, China, in 2019. She is working toward the M.S. degree in the field of signal and information processing from the School of Electrical Engineering, Southwest Jiaotong University, Chengdu, China. Her current research interests include active noise control, nonlinear signal processing, and adaptive filtering algorithms.